\documentclass[preprintnumbers,amsmath,amssymb,nofootinbib]{revtex4}

\usepackage{graphicx}
\usepackage{dcolumn}
\usepackage{bm}
\usepackage{float}
\usepackage[caption = false]{subfig}

\newcommand\ee{\end{equation}}
\newcommand\be{\begin{equation}}
\newcommand\eea{\end{eqnarray}}
\newcommand\bea{\begin{eqnarray}}

\begin{document}


\title{Bound states and the classical double copy}

\author{Walter D. Goldberger}
\affiliation{Physics Department, Yale University, New Haven, CT 06520, USA}
\author{Alexander K. Ridgway}
\affiliation{Walter Burke Institute for Theoretical Physics, California Institute of Technology, Pasadena, CA 91125, USA}

\date{\today}

\begin{abstract}

We extend the perturbative classical double copy to the analysis of bound systems.  We first obtain the leading order perturbative gluon radiation field sourced by a system of interacting color charges in arbitrary time dependent orbits, and test its validity by taking relativistic bremsstrahlung and non-relativistic bound state limits.   By generalizing the color to kinematic replacement rules recently used in the context of classical bremsstrahlung, we map the gluon emission amplitude to the radiation fields of dilaton gravity sourced by interacting particles in generic (self-consistent) orbits.   As an application, we reproduce the leading post-Newtonian radiation fields and energy flux for point masses in non-relativistic orbits from the double copy of gauge theory.

\end{abstract}

\maketitle

\section{Introduction}

The possibility that perturbative gravity is somehow the square of gauge theory was first concretely realized in the context of string theory, by the results of Kawai, Lewellen and Tye~\cite{KLT} (KLT), which formulated certain factorization identities relating $S$-matrix elements in closed and open string sectors.   In the limit of infinite string tension, these KLT relations connect tree-level gluon scattering amplitudes to corresponding ones involving graviton external states.   More recently, Bern, Carrasco and Johansson (BCJ) identified a more general perturbative ``duality" between scattering amplitudes of Yang-Mills and gravity theories~\cite{BCJ1,BCJ2,BCJ3}, which includes, a special case the earlier results of~\cite{KLT}.   Roughly speaking, BCJ duality states that once a Yang-Mills scattering amplitude is written in a certain canonical form, the corresponding gravity amplitude can be obtained by performing a simple set of color to kinematic replacements.  Although in field theory this correspondence has only been established at tree-level~\cite{BCJ2}, there is evidence of its validity at loop order~\cite{BCJ3}.  See~\cite{Review} for a recent review and a comprehensive list of references.

If, as implied by BCJ duality, gravitons can be interpreted as a ``double copy'' of gluons, it becomes natural to ask if other observables exhibit analogous structure.   Of particular interest is the question of whether the computational challenge of attempting to solve Einstein's equations, even perturbatively, can be sidestepped by applying a classical version of the double copy to analogous but relatively simpler solutions in Yang-Mills theory.   Investigation into this classical double copy was initiated in ref.~\cite{KS1} and further pursued in~\cite{KS2,KS3,KSothers} in the context of Kerr-Schild solutions of pure general relativity.  Recently, we~\cite{Goldberger:2016iau} showed that the classical double copy can be applied to the analysis of classical radiation fields and time-dependent sources.  Specifically, we found that the leading order\footnote{The perturbative double copy for classical static sources beyond leading order in perturbation theory was considered more recently, in~\cite{Luna:2016hge}.} classical bremsstrahlung radiation fields in Yang-Mills theory and dilaton gravity theory are related by a set of color to kinematic replacement rules, similar to those used in the context of scattering amplitudes.  This implies that, in the context of bremsstrahlung, all physical quantities of interest measured at asymptotic infinity predicted by gravity-dilaton theory can be derived from the computationally simpler vertices of gauge theory.

In the classical bremsstrahlung configurations considered in~\cite{Goldberger:2016iau}, the sources on the gauge theory side correspond to point color charges that start out widely separated, and therefore non-interacting in the far past.   This setup is sufficiently similar to the case of a five-point tree level scattering amplitude in the context of BCJ duality~\cite{Johansson:2014zca} that it is perhaps not unexpected that a consistent gravitational double copy holds in the classical limit as well.    See~\cite{Luna:2017dtq} for recent discussion.    In this paper, we extend the results of~\cite{Goldberger:2016iau} to the case of particles in general time-dependent orbits.   Our interest is primarily in the case of non-relativistic bound orbits, with the eventual goal of making contact with post-Newtonian compact binary inspirals~\cite{TheLIGOScientific:2017qsa}.    From a more formal perspective, such orbital configurations correspond to highly excited bound state poles in $S$-matrix elements, so that a simple tree-level interpretation of the radiation field and its double copy would be inadequate.   As such, these bound systems provide a new test of the classical double copy beyond fixed order perturbation theory in powers of the gauge coupling.

In sec.~\ref{Yang-Mills section}, we consider configurations of classical color charges in general time-dependent orbits, and solve the Yang-Mills equations to second order in a perturbative expansion.    The radiation field obtained in this way is then mapped onto an effective energy-momentum tensor ${\tilde T}^{\mu\nu}$ in dilaton gravity via a set of time-dependent color to kinematics transformation rules that we define in sec.~\ref{Double Copy}, generalizing those used in~\cite{Goldberger:2016iau}.   These transformation rules also map the equations of motion of the color charges to those of the corresponding point sources in the gravity theory.    The resulting double copy radiation fields are consistent with the relativistic bremsstrahlung solutions found earlier, but apply as well to more general orbital configurations.   As a special case, we work out in detail the case of non-relativistic configurations, finding agreement between the double copy prediction and earlier results in the literature~\cite{will,Damour} on tests of scalar-tensor theories of gravity.    Throughout, we work in $d=4$ spacetime dimensions, but our results generalize trivially to any $d$.

\section{Classical Yang-Mills radiation field for generic orbits}
\label{Yang-Mills section}

Ref.~\cite{Goldberger:2016iau} computed the long distance radiation field from a system of interacting color charges in classical Yang-Mills theory, for the case of orbits that are unbound, with the particles coming in from spatial infinity.    Our goal in this section is to show that the same methods can be applied to a much broader class of time-dependent orbits, including the case of bound systems in quasi-periodic trajectories.   We begin by reviewing the setup of~\cite{Goldberger:2016iau}.

By a color charge, we mean a point particle moving along a spacetime trajectory $x^{\mu}(\tau)$ that carries a degree of freedom $c^a(\tau)$ transforming locally (at $x^\mu(\tau)$) in the adjoint representation.    A concrete model for an ensemble of such particles is the worldline Lagrangian
\begin{align}
\label{worldline action}
S_{pp} &= -\sum\limits_{\alpha}m_{\alpha}\int d\tau + \sum\limits_{\alpha}\int dx^\mu \psi_{\alpha}^{\dagger} iD_\mu \psi_{\alpha}(\tau)
\end{align}
where the quantity $\psi(\tau)$ is a degree of freedom that transforms linearly in some representation of the gauge group\footnote{Our conventions are $D_\mu = \partial_\mu - i g A^a_\mu T^a$.}.    In terms of this variable, the color charge is then $c^a(\tau) = \psi^\dagger T^a \psi$.     For a fixed background gauge field, the equations of motion are expressible in terms of these adjoint color charges 
\begin{align}
\frac{dp^{\mu}_{\alpha}}{d\tau} &= - g c^{a}_{\alpha}(\tau) G^{a\mu}{}_{\nu}(x_{\alpha} )v^{\nu}_{\alpha}(\tau) \label{trajectory EOM},\\
\frac{dc_{\alpha}^{a}}{d\tau} &= -g f^{abc}v_{\alpha}^{\mu}(\tau)A^{b}_{\mu}(x_{\alpha}) c^{c}_{\alpha}(\tau) \label{color EOM}.
\end{align}
In fact, these equations of motion are independent of the specific model Lagrangian, in that, for dynamical gauge fields, they follow entirely from the Yang-Mills equation
\begin{align}
D_{\nu}G^{a\nu\mu}(x) = -g J^\mu_a(x) = - {\delta \over \delta A_\mu^a(x)} S_{pp},
\label{gluon EOM},
\end{align}
with field strength $-i g G^a_{\mu\nu} = [D_\mu,D_\nu]^a$, together with covariant conservation of the source current, $D_\mu J^\mu_a=0$, and conservation of stress-energy, $\partial_\nu\left(T^{\mu\nu}_{YM}+T^{\mu\nu}_{pp}\right)=0.$

Our interest here is in configurations of color charges that remain sufficiently far apart that the dynamics is perturbative over macroscopic time scales.    We consider particles separated by a typical distance scale $r$, and carrying energy $E\gtrsim m$.    The classical limit then corresponds to orbital angular momentum $L\sim E r\gg 1$ (we use units with $\hbar =1$).  In terms of $\alpha_s={g^2\over 4\pi}\ll 1$ there are two types of perturbative corrections in the regime $L\gg 1$.  Specifically, we are performing a double expansion in $\alpha_s c_a^2/L\ll 1,$ which controls the size of corrections to the orbital and color time evolution, as well as a parameter $\alpha_s c_a \ll 1$ that counts insertions of the classical gauge self-interactions.     This latter quantity can only dominate the parameter $\alpha_s \ll 1$ that counts quantum loop effects if the color charges $c_a$ are parametrically large in units of $\hbar$.    In fact, it was found in~\cite{Goldberger:2016iau} that a consistent double copy of radiating solutions requires the charges to be in a regime in which they scale as $c^a\sim L\gg 1$.  In this case both expansions coincide, and 
\begin{align}
\epsilon_{YM} =\alpha_s c^a \ll 1
\end{align}
is the small quantity that determines the size of perturbative corrections for generic kinematical configurations.    (Another scale in this problem is the typical frequency $\omega$ of emitted radiation.   We specialize to the case $\omega r\ll 1$ later on, but for now we keep it generic).

To set up the perturbative expansion, it is convenient to work in Feynman gauge, $\partial_\mu A^\mu_a=0$, in which the gluon equations of motion can be recast in the form
\begin{align}
\label{gluon EOM 2}
\Box A^{a\mu}(x) = -g\tilde{J}^{a\mu}(x),
\end{align}
where we have defined an ``effective source current" $\tilde{J}^{a\mu}(x)$ which is conserved, $\partial_\mu {\tilde J}^\mu_a=0,$ and sourced by both the particles and the gauge field itself, 
\begin{align}
\label{effective source current}
 \tilde{J}^{\mu}_a(x) =J^\mu_a(x) + f^{abc}A_{\nu}^{b}(x)\left(\partial^{\nu}A^{c\mu}(x) - G^{c\mu\nu}(x) \right).
\end{align}
This current contains all the physical information that is accessible to observers at asymptotic infinity.  For example, the radiation field at retarded time $t$ is given by  
\begin{align}
\label{YM radiation}
\lim_{r\rightarrow \infty}A^{\mu}_a(x) = -\frac{g}{4\pi r}\int \frac{d\omega}{2\pi}e^{-i\omega t}{\tilde J}^{\mu}_a(q),
\end{align}
where on the RHS, ${\tilde J}^{\mu}_a(q) = \int d^4 x e^{i q\cdot x} {\tilde J}^{\mu}_a(x)$ is evaluated at the on-shell four-momentum $q^\mu = \omega \left(1,{\vec{n}}\right)$, where ${\vec {n}}={{\vec x}\over |{\vec x}|}$ is the unit vector that points from the source to the detector.   This quantity also encodes the asymptotic distribution of energy-momentum and color radiated out to infinity, which can be expressed in terms of weighted integrals of the square of the polarized ``amplitude''
\begin{align}
\label{gluon emission amplitude}
{\cal A}^a(q) = g \epsilon_{\mu}\tilde{J}^{a\mu}(q),
\end{align}
with $q^2=0$ and $q\cdot \epsilon(q)=0$ both on-shell.   

\begin{figure}
\centering
\includegraphics[scale=0.3]{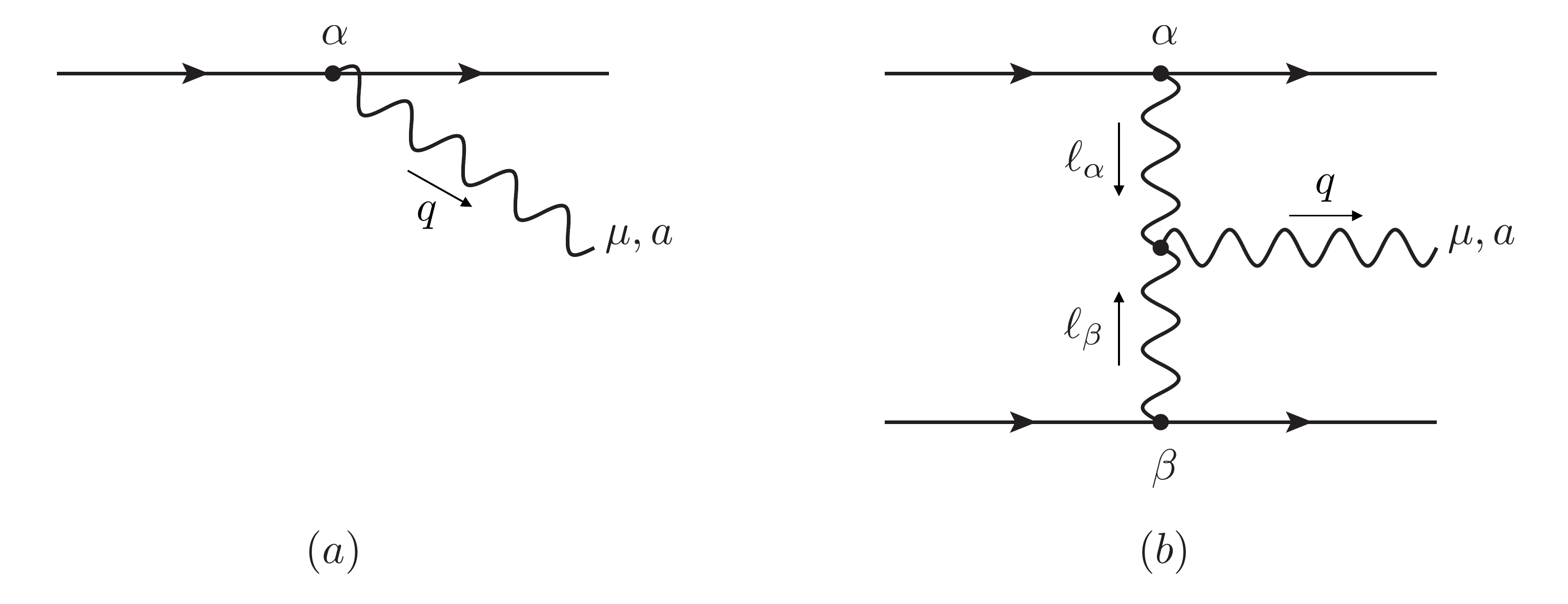}
\caption{Leading order Feynman diagrams for the perturbative expansion of ${\tilde J}^\mu_a(k)$.\label{fig:gluon1pt}}    
\end{figure}

The perturbative calculation of $\tilde{J}^{\mu}_a(x)$ can be expressed in terms of diagrams constructed from standard Yang-Mills bulk vertices and propagators, as well as insertions of the point particle currents.   Up to order ${\cal O}(g^2)$ beyond leading order, the relevant diagrams are shown in Fig.~\ref{fig:gluon1pt}.   We now compute these diagrams for arbitrary time-dependent trajectories $(x^\mu_\alpha,c^a_\alpha)(\tau)$, generalizing the results in~\cite{Goldberger:2016iau}.     The contribution from Fig.~\ref{fig:gluon1pt}(a) to the emission amplitude is simply given by 
\begin{eqnarray}
\label{worldline contribution}
\tilde{J}^{a\mu}(q)\bigg\rvert_{\rm worldline} &=&   \sum\limits_{\alpha}\int dx^\mu(\tau)_{\alpha} c^{a}_{\alpha}(\tau)e^{iq\cdot x_{\alpha}(\tau)}
={}  \sum_\alpha \int d\tau e^{iq\cdot x_{\alpha}}\frac{i}{q\cdot v_{\alpha}}\left[c^a_\alpha \left\{\dot{v}^{\mu}_{\alpha}- \frac{q\cdot\dot{v}_{\alpha}}{q\cdot v_{\alpha}} v_\alpha^\mu\right\} +  v^{\mu}_{\alpha}\dot{c}^{a}_{\alpha}\right]
\end{eqnarray}
where we have introduced the shorthand notation $x_\alpha = x^\mu(\tau_\alpha)$, $v^\mu_\alpha = {\dot x}^\mu_\alpha(\tau_\alpha)$, $c_\alpha = c_\alpha(\tau_\alpha)$ (we parametrize the worldlines by proper time $\tau$).   Note that to obtain the second equality we have performed integration by parts to put it in a form that will be convenient below.   The contribution from the cubic vertex, Fig.~\ref{fig:gluon1pt}(b) is 
\begin{eqnarray}
\label{cubic contribution 2}
\tilde{J}^{a\mu}(q)\bigg\rvert_{\rm cubic} &=& {i}g^2 \sum_{\alpha,\beta} \int d\tau_\alpha d\tau_\beta d\mu_{\alpha\beta}(q)  f^{abc} c^b_\alpha c^c_\beta \left[{1\over 2} (v_\alpha\cdot v_\beta)(\ell_\beta-\ell_\alpha)^\mu + (q\cdot v_\beta) v^\mu_\alpha- (q\cdot v_\alpha) v^\mu_\beta \right], 
\end{eqnarray}
where we have neglected terms which, after integration by parts, involve time derivatives on the worldline degrees of freedom and therefore subleading in powers of the gauge coupling (see Eqs.~(\ref{spatial trajectory}), (\ref{color trajectory}) below).    The momentum integral measure is defined as
\begin{equation}
d\mu_{\alpha\beta}(q) = \left[{d^4 \ell_\alpha\over (2\pi)^4}  {e^{i\ell_\alpha\cdot x_\alpha}\over \ell_\alpha^2}\right] \left[{d^4 \ell_\beta\over (2\pi)^4}  {e^{i\ell_\beta\cdot x_\beta}\over \ell_\beta^2}\right] (2\pi)^4\delta(\ell_\alpha+\ell_\beta -q)
\end{equation} 
with poles at $\ell_\alpha^2=0$ and $\ell_\beta^2=0$ corresponding to the propagators in Fig~\ref{fig:gluon1pt}(b).  It is understood that propagators obey retarded boundary conditions, $1/k^{2} = 1/[(k^{0} + i\epsilon)^{2} - \vec{k}^{2}]$ and $1/q\cdot v=1/(q\cdot v+i\epsilon)$, as is appropriate for classical observables.

These expressions can only yield consistent solutions that obey the Ward identity $q_\mu \tilde{J}^{a\mu}(q)=0$ if the color charges satisfy the classical equations of motion.   To leading order in perturbation theory, the time-dependent variables $(x^\mu_\alpha,c^a_\alpha)$ source a classical gauge field given by
\begin{align}
\label{LO potential field}
A^{a \mu}(x) = g \int {d^4 \ell\over (2\pi)^4} {e^{-i \ell\cdot x}\over \ell^2}  J^{a \mu}(\ell) =  g \sum_\alpha \int d\tau   {d^4 \ell\over (2\pi)^4}  {e^{-i \ell\cdot (x-x_\alpha)}\over \ell^2}  c^a_\alpha(\tau) v^\mu_\alpha(\tau).
\end{align}	
Plugging Eq.~(\ref{LO potential field}) into Eq.~(\ref{trajectory EOM}) and Eq.~(\ref{color EOM}) yields the equations of motion of the color charges at leading order,
\begin{eqnarray}
\label{spatial trajectory}
m_\alpha {\dot v}_\alpha &=& i g^2 \sum_\beta \int d\tau_\beta \frac{d^{4}\ell}{(2\pi)^{4}} {e^{-i\ell\cdot x_{\alpha\beta}} \over\ell^2} (c_\alpha\cdot c_\beta) \left[(v_\alpha\cdot v_\beta) \ell^\mu - (\ell\cdot v_\alpha) v^\mu_\beta\right],\\
\label{color trajectory}
{\dot c}_\alpha  &=& g^2 \sum_\beta \int d\tau_\beta \frac{d^{4}\ell}{(2\pi)^{4}} {e^{-i\ell\cdot x_{\alpha\beta}} \over\ell^2} f^{abc} c^b_\alpha c^c_\beta (v_\alpha\cdot v_\beta)
\end{eqnarray}
where $x_{\alpha\beta} = x_\alpha-x_\beta$.  Inserting (\ref{spatial trajectory}) and (\ref{color trajectory}) into Eq.~(\ref{worldline contribution}) and adding the result to Eq.~(\ref{cubic contribution 2}) then gives the total current at ${\cal O}(\alpha_s)$:
\begin{equation}
\label{eq:Jg2}
\tilde{J}^{a\mu}(q) =  g^2 \sum_{\alpha,\beta}\int {d\tau_\alpha d\tau_\beta}  d\mu_{\alpha\beta}(q)\left[i f^{abc} c_\alpha^b  c_\beta^c  {\cal A}^\mu_{adj} + (c_\alpha\cdot c_\beta) {c^a_\alpha\over m_\alpha} {\cal A}^\mu_{s}\right],
\end{equation}
where the amplitudes ${\cal A}^\mu_{adj,s}$ are given by
\begin{equation}
{\cal A}^\mu_{adj} =  (v_\alpha\cdot v_\beta)\left[{1\over 2} (\ell_\beta-\ell_\alpha)^\mu +{\ell_\alpha^2\over q\cdot v_\alpha} v_\alpha^\mu \right]+ (q\cdot v_\beta) v^\mu_\alpha - (q\cdot v_\alpha) v^\mu_\beta,
\end{equation}
and 
\begin{equation}
{\cal A}^\mu_s = -{\ell_\alpha^2\over q\cdot v_\alpha} \left[(v_\alpha\cdot v_\beta) \left(\ell^\mu_\beta -  {q\cdot\ell_\beta\over q\cdot v_\alpha} v^\mu_\alpha\right) -
(q\cdot v_\alpha) v^\mu_\beta + (q\cdot v_\beta) v^\mu_\alpha\right].
\end{equation}
It is manifest from the form of this equation that $q_\mu {\tilde J}^\mu_a(q)=0$.  Note $v^\mu_\alpha(\tau)$ and $c^a_\alpha(\tau)$ satisfy the equations of motion but are otherwise arbitrary.

\subsection{Consistency checks}

The result obtained in Eq.~(\ref{eq:Jg2}) holds for general orbits that obey the leading order equations of motion~(\ref{spatial trajectory}), (\ref{color trajectory}).   It contains as a special case the result of ref.~\cite{Goldberger:2016iau}, which considered classical scattering solutions, with orbits that asymptote to
\begin{align}
&x^\mu_\alpha(\tau\rightarrow-\infty) \rightarrow b^\mu_\alpha + v^\mu_\alpha \tau,\cr
&c^a(\tau\rightarrow-\infty)\rightarrow c_\alpha^a
\end{align}
with constant $b^\mu_\alpha$, $v^\mu_\alpha$, and $c^{a}_\alpha$ in the far past.   Indeed, simply inserting  constant orbital and color parameters into Eq.~(\ref{eq:Jg2}) exactly matches the solution found in~\cite{Goldberger:2016iau}.

We now verify that the general formula Eq.~(\ref{eq:Jg2}) also reproduces the correct non-relativistic limit.    Consider a system of particles in non-relativistic orbits (bound or unbound).  In this limit, the virial theorem implies that the typical velocity $v$ and orbital radius $r$ are related by 
\begin{equation}
 m v^2 \sim {\alpha_s c_a^2\over r},  
\end{equation}
and thus for the scaling $c^a\sim L=mvr$ relevant to the double copy (see below), the expansion parameter is $v \sim \alpha_s c_a\ll 1$ in agreement with the discussion in the section above.   In this limit the typical frequency of the orbits is $\omega_{orb}\sim v/r$ is the same as the that of the color factors, since  ${{\dot c_a}\over c_a} \sim \omega_c \sim \alpha_s c_a/r\sim \omega_{orb}$.

Working to leading order in $v^2\ll 1$, we can choose a Lorentz frame in which the particle trajectories take the form   
\begin{equation}
v^\mu_\alpha(\tau) =\left(1,{\vec v}_\alpha\right) + {\cal O}(v^2),
\end{equation}
where to the order we work to, ${\vec v}_\alpha = d{\vec x}_\alpha/dx^0$ is the ordinary three-velocity, with $|{\vec v}_\alpha|\ll 1$.   For such orbital configurations, the frequency of emitted radiation is parametrically of the same order as the orbital frequency, and the components of the outgoing gluon momentum obeys the scaling rule $q^\mu \sim (v/r,v/r)$.   On the other hand, the momentum (potential) exchange between the particles is typically off-shell with components that scale as $\ell^\mu\sim (v/r,1/r)$.   In this case, we may expand our general result in powers of ${\vec q}\cdot {\vec x}_\alpha\sim{\cal O}(v)$ (the multipole expansion) or powers of $\ell^0/|{\vec \ell}|\sim {\cal O}(v)$ (retardation effects), but must treat ${\vec \ell}\cdot {\vec x}_\alpha\sim {\cal O}(1)$ non-perturbatively.   In this limit, the equations of motion~(\ref{spatial trajectory}) and~(\ref{color trajectory}) reduce to
\begin{equation}
\label{eq:nrymv}
m_\alpha {\dot{\vec v}}_\alpha(t) = -ig^2 \sum_\beta (c_\alpha\cdot c_\beta) \int {d^3{\vec\ell}\over (2\pi)^3} {e^{i{\vec\ell}\cdot {\vec x}_{\alpha\beta}(t)} \over \vec\ell^2} {\vec\ell} = \alpha_s \sum_\beta  {
({c_\alpha\cdot c_\beta}){\vec x}_{\alpha\beta} \over |{\vec x}_{\alpha\beta}|^3},
\end{equation}
and
\begin{equation}
\label{eq:nrymc}
{\dot c}_\alpha(t) = -g^2 \sum_\beta f^{abc} c^b_\alpha c^c_\beta \int {d^3{\vec\ell}\over (2\pi)^3} {e^{i{\vec\ell}\cdot {\vec x}_{\alpha\beta}(t)} \over \vec\ell^2} = -\alpha_s\sum_\beta   {f^{abc} c^b_\alpha c^c_\beta\over   |{\vec x}_{\alpha\beta}|},
\end{equation}
respectively.    

Given the scaling of radiation and potential momenta, it is easy to see that in the non-relativistic limit with $c^a\sim L$, the contribution from ${\cal A}_s^\mu$ in Eq.~(\ref{eq:Jg2}) is comparable to that of ${\cal A}_{adj}^\mu$.    Our solution for ${\tilde J}^\mu_a(q)$ at ${\cal O}(\alpha_s)$ is manifestly gauge invariant,  but it is convenient to calculate the radiation field in a gauge where the gluon polarization is purely spatial, in which case the relevant part Eq.~(\ref{eq:Jg2}) simplifies to
\begin{eqnarray}
\nonumber
{\tilde J}^i_a(q) &=& ig^2 \sum_{\alpha,\beta} \int dt e^{i\omega t}  \int {d^3{\vec \ell}\over (2\pi)^3}  {e^{i{\vec\ell}\cdot {\vec x}_{\alpha\beta}(t)}\over{\vec \ell}^4} \left[{{\vec\ell^2}\over \omega+i\epsilon}\left(-i(c_{\alpha}\cdot c_{\beta}){c^a_\alpha\over m_\alpha}{ \ell}^i - f^{abc} c^b_\alpha c^c_\beta { v}^i_\alpha\right) 
+ f^{abc} c^b_\alpha c^c_\beta  {\ell}^i\right]\\
\nonumber\\
&= &i \int dt e^{i\omega t} \left[{1\over \omega+i\epsilon}\sum_{\alpha}\left(c^a_\alpha {\dot { v}^i}_\alpha + {\dot c}_\alpha { v}^i_\alpha\right)-{g^2\over 2}\sum_{\alpha\beta}  f^{abc} c^b_\alpha c^c_\beta\int {d^3{\vec \ell}\over (2\pi)^3} e^{i{\vec\ell}\cdot {\vec x}_{\alpha\beta}(t)}
 {\partial\over \partial{\ell^i}} \left({1\over{\vec \ell}^2}\right)\right]   
 \end{eqnarray}
at leading order in the NR limit.   The last term can be written as
\begin{equation}
i{\alpha_s\over 2}\int dt e^{i\omega t} \sum_{\alpha\beta}  f^{abc} c^b_\alpha c^c_\beta {{\vec x}_{\alpha\beta}\over |\vec x_{\alpha\beta}|} =  -i\int dt e^{i\omega t} \sum_\alpha {\dot c}^a_\alpha {\vec x}_\alpha ,
\end{equation}
so that after integration by parts on the first term,  the leading order non-relativistic current becomes
\begin{equation}
{\tilde J}_a^i(q) = \int dt e^{i\omega t} \sum_{\alpha}\left(c^a_\alpha {\dot {\vec x}}_\alpha  + {\dot c}^a_\alpha {\vec x}_\alpha\right)^i =  -i\omega \int dt e^{i\omega t}{ p}^i_a(t),
\end{equation}
where ${\vec p}_a(t) =  \sum_\alpha c^a_\alpha {\vec x}_\alpha$ is the net color electric dipole moment of the system of charges.    The emission amplitude ${\cal A}^a(q) = i\omega \epsilon_i {{\tilde J}}^i_a(q)$ is precisely what we would have obtained from a composite non-relativistic object that carries a time-dependent dipole ${\vec p}_a(t)$ along its orbit, and couples to the color electric field through a point-particle interaction that takes the form
\begin{equation}
S_{int} = -\int dt\, {\vec p}_a(t) \cdot {\vec E}^a(t,{\vec 0}),
\end{equation}
in the composite object's rest frame.   We therefore recover the expected answer to leading order in the velocity expansion~\cite{Grinstein:1997gv}.

\section{Double Copy}
\label{Double Copy}

We now generalize the classical double copy rules discussed in~\cite{Goldberger:2016iau} to the case of general time-dependent orbits, building on the result in Eq.~(\ref{eq:Jg2}).     By applying the formal substitution rules\footnote{The replacement rules listed in (\ref{masterpiece rules}) are slightly different than those used in \cite{Goldberger:2016iau}.  In particular, the structure constant and color charge mappings are different by factors of $i$.  Despite this, applying (\ref{masterpiece rules}) to the gluon emission amplitude due to bremsstrahlung will give the same result as given in \cite{Goldberger:2016iau} up to an irrelevant phase.}
\begin{eqnarray}
\label{masterpiece rules}
\nonumber
c^a_\alpha(\tau) &\rightarrow&  i m_\alpha v_\alpha^\nu(\tau),\\
f^{abc} c^a_\alpha c^b_\beta &\rightarrow& {1\over 2} m_\alpha m_\beta\left[(v_\alpha\cdot v_\beta) (\ell_\beta-\ell_\alpha)^\nu + v_\beta\cdot (\ell_\alpha+q) v_\alpha^\nu-  v_\alpha\cdot (\ell_\beta+q) v_\beta^\nu \right],\\
\nonumber
m_\alpha v^\mu_\alpha(\tau) &\rightarrow& m_\alpha v^\mu_\alpha(\tau),
\end{eqnarray}
and $g\rightarrow {1/2 m_{Pl}}$, to the the current ${\tilde J}^\mu_a(q)$, we obtain an object ${\tilde J}^{\mu}_a(q)\rightarrow i {\tilde T}^{\mu\nu}(q)$\footnote{In our expression for  ${\tilde T}^{\mu\nu}(q)$ have dropped terms subleading in $\epsilon_g$ that involve the accelerations ${\dot v}_\alpha$, and have added an ``improvement term,'' proportional to $q^\mu$, which does not affect physical quantities to write 
$$
{\hat {\cal A}}^\mu_s = -{\ell_\alpha^2\over q\cdot v_\alpha} \left[(v_\alpha\cdot v_\beta) \left({1\over 2}(\ell_\beta-\ell_\alpha)^\mu -  {q\cdot\ell_\beta\over q\cdot v_\alpha} v^\mu_\alpha\right) -
(v_\alpha\cdot q) v^\mu_\beta +(q\cdot v_\beta)v^\mu_\alpha\right].
$$}
with 
 \begin{eqnarray}
 \label{eq:Tmunu}
\nonumber
{\tilde T}^{\mu\nu}(q) &=& {1\over 4 m_{Pl}^2} \sum_{\alpha,\beta}m_\alpha m_\beta \int {d\tau_\alpha d\tau_\beta}  d\mu_{\alpha\beta}(q)\left[\left( {1\over 2} (v_\alpha\cdot v_\beta) (\ell_\beta-\ell_\alpha)^\nu + (v_\beta\cdot q) v_\alpha^\nu-  (v_\alpha\cdot q) v_\beta^\nu\right) {\cal A}^\mu_{adj}\right. \\
& &\hspace{7cm} {}\left. - (v_\alpha\cdot v_\beta) {v^\nu_\alpha} {\hat {\cal A}}^\mu_{s}\right],
\end{eqnarray}
which is symmetric and for on-shell momenta $q^2=0$ satisfies $q_\mu {\tilde T}^{\mu\nu}(q) =0$.  The on-shell tensor  ${\tilde T}^{\mu\nu}(q)$ defines a self-consistent perturbative solution in a theory of point sources coupled to gravity.    This solution correspond a system of weakly gravitating sources, which is true for generic kinematics whenever the expansion parameter $\epsilon_g\sim G_N E/r\ll 1$ (we define $G_N = 1/32 \pi m_{Pl}^2$).   Specifically, ${\tilde T}^{\mu\nu}(q)$ sources a helicity-2 radiation field given by
\begin{equation}
h_{\pm}(t,{\vec n}) = {4 G_N\over r} \int {d\omega\over 2\pi} e^{-i\omega t} \epsilon^{*\mu\nu}_{\pm}(q) {\tilde T}_{\mu\nu}(q),
\end{equation}
as well as a radiation scalar 
\begin{equation}
\phi(t,{\vec n}) =  {G_N\over r} \int {d\omega\over 2\pi} e^{-i\omega t} {\tilde T}^{\mu}{}_\mu(q).
\end{equation}
Given the analytic structure of the integrand in ${\tilde T}^{\mu\nu}(q)$, this theory of gravity must be local.  In particular, the residue of the double pole in the integrand at ${\ell}^2_\alpha={\ell}^2_\beta =0$ is analytic in kinematic variables, so encodes the local cubic interactions among the fields in the gravitational sector.

Indeed, for the case of inelastic scattering with constant velocities $v_\alpha$ (to leading order in $\epsilon_g$),  Eq.~(\ref{eq:Tmunu}) precisely matches the results of~\cite{Goldberger:2016iau}, which identified the double copy amplitude ${\tilde T}^{\mu\nu}(q)$ with relativistic bremsstrahlung scattering solutions of a particular dilaton gravity theory, whose action in $d=4$ is given by 
\begin{align}
\label{gravity dilaton}
S_{g} = -2m^{2}_{pl}\int d^{4}x\sqrt{-g}\left[R - 2g^{\mu\nu}\partial_{\mu}\phi\partial_{\nu}\phi \right] - \sum\limits_{\alpha}m_{\alpha}\int d\tau_{\alpha} e^{\phi}.
\end{align}
More generally, we now see that Eq.~(\ref{eq:Tmunu}) yields solutions of this theory with particles in general orbital configurations.

\subsection{Application to post-Newtonian systems}

As a test of the general result Eq.~(\ref{eq:Tmunu}), we consider particles in non-relativistic orbits, with velocities $v^\mu_\alpha=(1,{\vec v}_\alpha)+{\cal O}(v^2)$.    The integrals in Eq.~(\ref{eq:Tmunu}) are then dominated by momentum regions of potential exchange $\ell\sim (v/r,1/r)$ and radiation $q\sim(v/r,v/r)$ (the systematics and general power counting for this expansion are discussed in~\cite{Goldberger:2004jt}).     The orbital equation in gravity is obtained by applying the mapping defined in Eq.~(\ref{masterpiece rules}) to either the non-relativistic color or orbital equations in gauge theory, Eqs.~(\ref{eq:nrymv}),~(\ref{eq:nrymc}).   To lowest order in velocity this is
\begin{equation}
\label{eq:newt}
m_\alpha {\dot {\vec v}}_\alpha = {i\over 4 m_{Pl}^2} \sum_\beta  \int {d^3{\vec\ell}\over (2\pi)^3} e^{i{\vec\ell}\cdot {\vec x}_{\alpha\beta}} {\vec \ell\over {\vec\ell}^2} = - 2 G_N \sum_\beta {m_\alpha m_\beta {\vec x}_{\alpha\beta}\over |{\vec x}_{\alpha\beta}|^3},
\end{equation}
implying the conservation of energy $E={1\over 2}\sum_\alpha m_\alpha {\vec v}^2_\alpha - \sum_{\alpha\beta} G_N m_\alpha m_\beta/{|{\vec x}_{\alpha\beta}|}=\sum m_\alpha  \left({1\over 2} {\vec v}^2_\alpha + {\vec x}_\alpha \cdot \ddot{\vec x}_\alpha\right)$.    Note that by defining $G_N$ as in pure Einstein gravity,  Eq.~(\ref{eq:newt}) differs by a factor of two from Newton's law, reflecting the additional contribution of dilaton exchange to the potential.

We find it convenient to work in a gauge with purely spatial graviton polarizations, in which case the relevant terms at leading order in ${\tilde T}^{\mu\nu}(q)$ at order $v^2\sim \epsilon_g$ are
\begin{eqnarray}
\label{eq:PN}
\nonumber
{\tilde T}^{ij}(q) &=&  \sum_{\alpha,\beta} {m_\alpha m_\beta \over 4 m_{Pl}^2} \int dt e^{i\omega t}  \int {d^3{\vec \ell}\over (2\pi)^3} {e^{i{\vec\ell}\cdot {\vec x}_{\alpha\beta}(t)}\over{\vec \ell}^4} \left[ \ell^i \ell^j - {{\vec\ell}^2\over \omega+i\epsilon} (v_\alpha^i \ell^j + v_\alpha^j \ell^i)\right]\\
&=& \int dt e^{i\omega t} \sum_\alpha m_\alpha \left[{i\over \omega+i\epsilon} {d\over d t}\left(v_\alpha^i v_\alpha^j\right) + \sum_\beta {G_N m_\beta\over |{\vec x}_{\alpha\beta}|}\left(\delta^{ij} - {{ x}_{\alpha\beta}^i { x}_{\alpha\beta}^j\over |{\vec x}_{\alpha\beta}|^2}\right) \right].
\end{eqnarray}
Using the identity
\begin{equation}
\sum_{\alpha\beta}  {G_N m_\alpha m_\beta\over |{\vec x}_{\alpha\beta}|^3}  { x}_{\alpha\beta}^i { x}_{\alpha\beta}^j =\sum_\alpha m_\alpha x^i_\alpha \sum_\beta {G_N m_\beta { x}^j_{\alpha\beta}\over |{\vec x}_{\alpha\beta}|^3} + (i\leftrightarrow j) =-{1\over 2} \sum_\alpha m_\alpha \left({ x}_\alpha^i {\ddot { x}}_\alpha^j + (i\leftrightarrow j)\right),
\end{equation}
we can write ${\tilde T}^{ij}(q)$, after integration by parts
\begin{equation}
{\tilde T}^{ij}(q) =  \int dt e^{i\omega t} \sum_\alpha m_\alpha \left[{1\over 2} {d^2\over d t^2} { x}_\alpha^i { x}_\alpha^j - \delta^{ij} {\vec x}_\alpha\cdot \ddot{\vec x}_\alpha\right],
\end{equation}
This yields a canonically normalized graviton and scalar emission amplitude
\begin{eqnarray}
\label{eq:pngrav}
i{\cal A}_g(q) &=& {i\omega^2 \over   4 m_{Pl}} \int dt e^{i\omega t} \epsilon^*_{ij}(q) Q^{ij}(t),\\
\label{eq:pndil}
i{\cal A}_\phi(q) &=& {i\over 2\sqrt{2} m_{Pl}}\int dt e^{i\omega t} \left(\delta^{ij} - {{ q}^i { q}^j\over { q}^2}\right) {\tilde T}^{ij}(q) = {i\over 2\sqrt{2} m_{Pl}} \int dt e^{i\omega t} \sum_\alpha m_\alpha \left( {\vec v}_\alpha^2  - {\vec x}_\alpha\cdot \ddot{\vec x}_\alpha+{1\over 2} ({\vec x}_\alpha\cdot {\vec q})^2 \right).
\end{eqnarray}
with quadrupole moment $Q^{ij}=\sum_\alpha m_\alpha \left({ x}^i_\alpha  { x}^j_\alpha-{1\over 3}\delta^{ij} {\vec x}_\alpha^2\right)$.   Equivalently, the radiation fields at null infinity ($r\rightarrow\infty$ and retarded time $t$) are to leading post-Newtonian order,
\begin{eqnarray}
\label{eq:hrad}
h_{ij} (t,{\vec n}) &=& {2 G_N\over r} [{\ddot Q}_{ij}(t)]^{TT},\\
\label{eq:prad}
\phi(t,{\vec n})   &=& {G_N\over r} \sum_\alpha m_\alpha \left( {\vec v}_\alpha^2  - {\vec x}_\alpha\cdot \ddot{\vec x}_\alpha -{1\over 2} {d^2\over dt^2} ({\vec x}_\alpha\cdot {\vec n})^2\right),
\end{eqnarray}
where $TT$ denotes the transverse traceless part, using the projector $\delta_{ij}-n_i n_j$, ${\vec n}={\vec x}/|{\vec x}|$.   

This is the radiation pattern that is sourced by a composite particle at rest at the origin, whose stress tensor is of the form
\begin{equation}
{\tilde T}^{ij}(t,{\vec x})= \sum_\alpha m_\alpha \left[{1\over 2} {d^2\over d t^2} { x}_\alpha^i { x}_\alpha^j - \delta^{ij} {\vec x}_\alpha\cdot \ddot{\vec x}_\alpha\right]\delta^3({\vec x}).
\end{equation}
At the linear level, the interactions of this composite particle with gravity are given by a worldline Lagrangian of the form~\cite{Goldberger:2005cd}
\begin{equation}
\label{eq:gravints}
S_{int} =  \int dt Q_\phi(t) \phi(t,{\vec 0}) +{1\over 2} \int dt Q_{ij}(t) \left(E_{ij} - \partial_i \partial_j \phi\right)(t,{\vec 0})
\end{equation}
where the ``gravito-electric" field is related to the Weyl tensor, $E_{ij}=W_{0i0j}$, and the scalar monopole charge $Q_\phi(t)$ is proportional to the non-relativistic Lagrangian of the point-particles, 
\begin{equation}
Q_\phi(t) = {4\over 3} L = {2\over 3}\sum_\alpha m_\alpha {\vec v}^2_\alpha + {4\over 3}\sum_{\alpha\beta} {G_N m_\alpha m_\beta\over |{\vec x}_{\alpha\beta}|} = {2\over 3}\sum_\alpha m_\alpha \left({\vec v}_\alpha^2 - 2 {\vec x}_\alpha\cdot\ddot{\vec x}_\alpha\right).
\end{equation}
The scalar dipole moment is proportional to ${\vec P}=\sum_\alpha m_\alpha \dot{\vec x}_\alpha$, which is constant at this order in the velocity expansion so does not contribute to radiation.   Note that, on-shell, the linearized quadrupole interaction can also be expressed in terms of the Riemann tensor associated with ${\tilde g}_{\mu\nu}=e^{2\phi} g_{\mu\nu}$  simply as ${1\over 2} \int dt Q_{ij}(t) {\tilde R}_{0i0j}$.

It follows from Eq.~(\ref{eq:gravints}) that the total (time averaged) graviton energy flux is given by the standard quadrupole formula $P_g = {G_N\over 5}  \langle \dddot{Q}_{ij} \dddot{Q}_{ij} \rangle$, while in the scalar channel, the radiated power is given by 
\begin{equation}
P_\phi= {4 G_N} \left[\langle {\dot Q}_\phi^2\rangle + {1\over 30}\langle {\dddot Q}_{ij}^2\rangle\right].
\end{equation}    
Refs.~\cite{will,Damour} computed the radiation fields and energy fluxes in generalized scalar-tensor theories of gravity, which include as a special case the dilaton theory defined in Eq.~(\ref{gravity dilaton}).    We have verified both by direct calculation and by comparison with~\cite{will,Damour} that  Eqs.~(\ref{eq:hrad}),~(\ref{eq:prad}) are in agreement with the radiation fields of dilaton gravity (see for instance Eqs. (2.18), (2.19) of ref.~\cite{will}), providing a consistency check of the generalized double copy formula Eq.~(\ref{eq:Tmunu}).

\section{Conclusion}

We have extended the classical double copy to the analysis of gravitational radiation sourced by point sources in generic orbits.  In order to do this, we introduced a natural generalization of the color to kinematics replacements rules used in \cite{Goldberger:2016iau}, which maps the gluon field of a generic configuration of interacting color sources into a radiating solution in dilaton gravity coupled to point particles.   

For non-relativistic bound orbits, the double copy takes a particularly simple form.  The color dipole moment ${\vec p}_a$ of the gauge theory bound state maps onto a gravitational quadrupole moment $Q_{ij}$, which couples to the string frame Riemann tensor component ${\tilde R}_{0i0j}$, and to a scalar charge proportional to the gravitational Lagrangian of the particles, which measures the non-conservation of the dilatation current.   To make contact with sources in pure gravity, such as the inspiral events recently reported in~\cite{TheLIGOScientific:2017qsa}, a systematic way to remove the contributions due to dilaton interactions needs to be developed.   For progress in this direction see~\cite{Luna:2017dtq}.  It also remains to be seen if the classical double copy of radiation via the rules in Eq.~(\ref{masterpiece rules}) continues to hold beyond the leading order in perturbation theory.    Given the relative simplicity of the non-relativistic limit, the analysis of such higher order corrections to the double copy might be more transparent directly in terms of an effective Lagrangian with manifest velocity power counting~\cite{Goldberger:2004jt}, and with Feynman rules that are optimized~\cite{Kol:2007bc,Gilmore:2008gq} for post-Newtonian systems.  Finally, the effects of intrinsic spin in the gravitational dynamics~\cite{Porto:2005ac}, which are important for astrophysical sources, must be accounted for within the double copy.

\section{Acknowledgments}

AKR thanks Leo Stein and Mark Wise.  WG thanks W. Skiba and S. Prabhu for discussions at early stages of this work, and J. Li for crucial information.   This research was partially supported by Department of Energy grants DE-FG02-92ER-40704 (WG) and DE-SC0011632 (AKR),  and by the Gordon and Betty Moore Foundation through Grant No. 776 to the Caltech Moore Center for Theoretical Cosmology and Physics (AKR).

\end{document}